\pdfoutput=1
\documentclass[graybox]{svmult}
\usepackage{mathptmx}       
\usepackage{graphicx}        
\usepackage{amssymb}
\usepackage{amsmath} 
\begin{document}

\title*{Constructing the self-force}
\author{Eric Poisson}
\institute{Eric Poisson \at Department of Physics, University of
  Guelph, Guelph, Ontario, N1G 2W1 Canada. \email{poisson@physics.uoguelph.ca}}
\maketitle

\abstract{
I present an overview of the methods involved in the computation of
the scalar, electromagnetic, and gravitational self-forces acting on a
point particle moving in a curved spacetime. For simplicity, the focus
here will be on the scalar self-force. The lecture follows closely
my review article on this subject [E.\ Poisson, Living Rev.\
Relativity {\bf 7}, (2004), http://www.livingreviews.org/lrr-2004-6]. 
I begin with a review of geometrical elements (Synge's world function,
the parallel propagator). Next I introduce useful coordinate systems
(Fermi normal coordinates and retarded light-cone coordinates) in a
neighborhood of the particle's world line. I then present the wave
equation for a scalar field in curved spacetime and the equations of
motion for a particle endowed with a scalar charge. The wave equation
is solved by means of a Green's function, and the self-force is
constructed from the field gradient. Because the retarded field is
singular on the world line, the self-force must involve a regularized
version of the field gradient, and I describe how the regular piece of
the self-field can be identified. In the penultimate section of the
lecture I put the construction of the self-force on a sophisticated
axiomatic basis, and in the concluding section I explain how one can
do better by abandoning the dangerous fiction of a point particle.}

\section{Introduction} 

We consider a point particle moving on a world line $\gamma$ in a 
curved spacetime with metric $g_{\alpha\beta}$. The particle is either
endowed with a scalar charge $q$ or an electric charge $e$, and we
wish to calculate the effect of these charges on the motion of the
particle. This motion is not geodesic, because the (scalar or
electromagnetic) field created by the particle interacts with the
particle and causes it to accelerate. In flat spacetime this effect is
produced by a local distortion of the field lines associated with the
particle's acceleration. In curved spacetime there is no such local
distortion when the particle moves freely; what happens instead is
that the field interacts with the spacetime curvature and
back-scatters toward the particle. As far as the particle is
concerned, then, it interacts with an incoming wave, and the motion is
not geodesic.   

There is a gravitational analogue to these (scalar and
electromagnetic) situations: Even in the absence of charges $q$ and
$e$, we may wish to go beyond the test-mass description and consider 
the effect of the particle's mass $m$ on its motion. There are two
ways of describing this effect. We might say that the particle moves
on a {\it geodesic} in a {\it perturbed  spacetime} with metric
$g_{\alpha\beta} + h_{\alpha\beta}$.  Or we might say that the
particle moves on an {\it accelerated world line} in the 
{\it original spacetime} with metric $g_{\alpha\beta}$. It is useful
to keep both points of view active, and most researchers working in
this field go freely back and forth between these modes of
description. In the second view, the particle's acceleration is
associated with the perturbation $h_{\alpha\beta}$, which produces a
gravitational self-force acting on the particle.  

The scalar, electromagnetic, and gravitational self-force problems
share many physical and mathematical features. In each case a moving 
charge is accompanied by a field: $q$ produces a scalar field
$\Phi$, $e$ produces an electromagnetic field $A_\alpha$, and $m$
produces a gravitational perturbation $h_{\alpha\beta}$. In each case
the field satisfies a linear wave equation in the background
spacetime: $\Phi$ satisfies a scalar wave equation, $A_\alpha$ a
vectorial equation derived from Maxwell's equations, and 
$h_{\alpha\beta}$  a tensorial equation derived from the linearized
Einstein equations. And in each case the self-force is equal to the
gradient of the field evaluated on the particle's world line.  

This last observation reveals the problematic nature of this
investigation. It is straightforward enough to calculate the scalar
field $\Phi$, the vector potential $A_\alpha$, and the gravitational 
perturbation $h_{\alpha\beta}$ at a distance from the particle. But
because the particle is pointlike, these quantities diverge on the
world line, and derivatives of these quantities are even more
singular. How is one supposed to deal with these singular
expressions and extract from them the finite pieces that produce a
well-defined effect, namely the self-force acting on the particle? 

The purpose of this lecture is to offer some elements of answer to
this question. My focus will be on the technical aspects of the
problem, which I will try to describe without going overboard with 
derivations and mathematical precision. The preceding lecture by Bob
Wald offers more on the conceptual aspects of the self-force, and
other contributions describe ways of computing the self-force and its
consequences. This lecture can be considered to be a light
introduction to my massive review article  published in 
{\it Living Reviews in Relativity} \cite{poisson:04b}, to which I will
frequently refer (as LRR); there the reader will find all the gory
details of all the derivations omitted in the lecture.  

For simplicity I will focus on the simplest exemplar of a self-force:
the {\it scalar self-force} produced by a scalar field $\Phi$ on a
scalar charge $q$. Understanding the details of this construction is
a first step toward understanding the nature of the electromagnetic 
and gravitational self-forces; the computations involved are simpler,
but the conceptual basis is essentially the same.   

The self-force has a long history in theoretical physics, which is
nicely summarized in a book by Herbert Spohn \cite{spohn:08}. The
standard reference for the electromagnetic self-force in flat
spacetime is Dirac's famous 1938 paper \cite{dirac:38}. Dirac's
construction was generalized to curved spacetime in 1960 by DeWitt and
Brehme \cite{dewitt-brehme:60}; a technical error in their work was
corrected by Hobbs \cite{hobbs:68}. The gravitational self-force was
first computed in 1997 by Mino, Sasaki, and Tanaka
\cite{mino-etal:97}; a more direct derivation (based on an axiomatic
approach) was later provided by Quinn and Wald
\cite{quinn-wald:97}. Finally, the scalar self-force --- the main
topic of this lecture --- was first constructed in 2000 by Ted Quinn 
\cite{quinn:00}. 

\section{Geometric elements} 

The construction of the self-force would be impossible without the
introduction of geometric tools that were first fashioned by Synge
\cite{synge:60} and independently by DeWitt and Brehme
\cite{dewitt-brehme:60}. In this section I introduce the
world function $\sigma(x,x')$ and the parallel propagator 
$g^\alpha_{\ \alpha'}(x,x')$. 

\begin{figure} 
\includegraphics[width=0.6\linewidth]{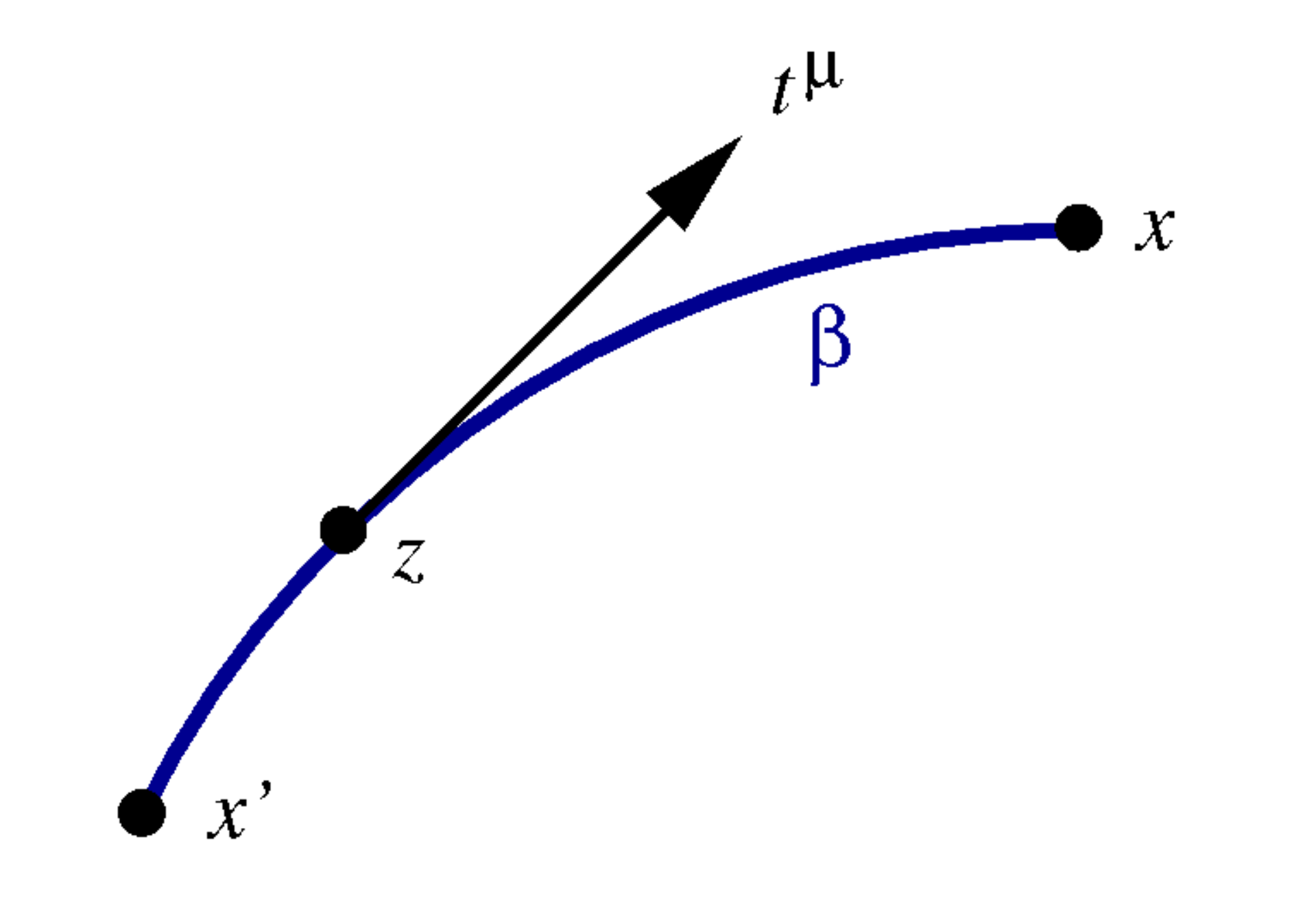}
\caption{Geodesic segment $\beta$ between the spacetime points $x'$ and
$x$. The vector $t^\mu$ is tangent to this geodesic.}   
\end{figure} 

Let $x$ and $x'$ be two points in spacetime, and let us assume that
they are sufficiently close that there is a unique geodesic segment
$\beta$ linking them. The segment is described by the parametric
relations $z^\mu(\lambda)$, in which $\lambda$ is an affine parameter
that runs from $0$ to $1$; we have that $z(0) = x'$ and $z(1) =
x$. The vector  $t^\mu := dz^\mu/d\lambda$ is tangent to $\beta$. 

The {\it world function} is defined by 
\begin{equation} 
\sigma(x,x') := \frac{1}{2} \int_0^1 g_{\mu\nu}(z) t^\mu t^\nu\,
d\lambda .
\end{equation} 
It is numerically equal to half the squared geodesic distance between
$x$ and $x'$. When $\sigma < 0$ the separation between $x$ and $x'$ is
timelike, and when $\sigma > 0$ it is spacelike. The equation
$\sigma(x,x') = 0$ describes the light cones of each point; if $x'$ is
kept fixed then $\sigma(x) = 0$ describes the past and future light
cones of $x'$; if instead $x$ is kept fixed then $\sigma(x') = 0$
describes the past and future light cones of $x$. 

The world function can be differentiated with respect to each
argument. It can be shown (LRR Sec.~2.1.2) that 
\begin{equation} 
-\sigma^{\alpha'} := -\nabla^{\alpha'} \sigma(x,x') 
\end{equation} 
is a vector at $x'$ that is proportional to $t^\mu$ evaluated at that
point. I use the convention that primed indices refer to the point
$x'$, while unprimed indices refer to $x$. It can also be shown that
its length is given by  
\begin{equation} 
g_{\alpha'\beta'} \sigma^{\alpha'} \sigma^{\beta'} = 2\sigma; 
\end{equation} 
the length of $\sigma^{\alpha'}$ therefore measures the geodesic 
distance between the two points. Because the vector points from $x'$
to $x$, we have a covariant notion of a displacement vector between
the two points. 

The {\it parallel propagator} takes a vector $A^{\alpha'}$ at $x'$ and
moves it to $x$ by parallel transport on the geodesic segment
$\beta$. We express this operation as 
\begin{equation} 
A^\alpha(x) = g^\alpha_{\ \alpha'}(x,x') A^{\alpha'}(x'),  
\end{equation}   
in which $A^\alpha$ is the resulting vector at $x$. The operation is
easily generalized to dual vectors and other types of
tensors (LRR Sec.~2.3).  

The world function and the parallel propagator can be employed in the
construction of a Taylor expansion of a tensor about a reference point
$x'$. Suppose that we have a tensor field $A^{\alpha\beta}(x)$ and
that we wish to express it as an expansion in powers of the
displacement away from $x'$. The role of the deviation vector is
played by $-\sigma^{\alpha'}$, and the expansion coefficients will be
ordinary tensors at $x'$. We might write something like 
\[
A^{\alpha'\beta'} 
+ A^{\alpha'\beta'}_{\ \ \ \ \gamma'} (-\sigma^{\gamma'}) 
+ \frac{1}{2} A^{\alpha'\beta'}_{\ \ \ \ \gamma'\delta'} 
  (-\sigma^{\gamma'})(-\sigma^{\delta'}) + \cdots, 
\]
but this defines a tensor at $x'$, not $x$. To get a proper
expression for $A^{\alpha\beta}(x)$ we must also involve the parallel
propagator, and we write 
\begin{equation} 
A^{\alpha\beta} = g^{\alpha}_{\ \alpha'} g^{\beta}_{\ \beta'} 
\biggl[ A^{\alpha'\beta'} 
- A^{\alpha'\beta'}_{\ \ \ \ \gamma'} \sigma^{\gamma'} 
+ \frac{1}{2} A^{\alpha'\beta'}_{\ \ \ \ \gamma'\delta'} 
  \sigma^{\gamma'} \sigma^{\delta'} + \cdots \biggr]. 
\end{equation} 
Having postulated this form for the expansion, the expansion
coefficients $A^{\alpha'\beta'}$, 
$A^{\alpha'\beta'}_{\ \ \ \ \gamma'}$, and so on can be computed by
repeatedly differentiating the tensor field and evaluating the results
in the limit $x \to x'$ (see LRR Sec.~2.4). For example,
$A^{\alpha'\beta'} = \lim\, A^{\alpha\beta}$, as we might expect.    

\section{Coordinate systems} 

Self-force computations are best carried out using covariant
methods. It is convenient, however, to display the results in a
coordinate system that is well suited to the description of a
neighbourhood of the world line $\gamma$. In this transcription it is
advantageous to keep the coordinates in a close correspondence with
the geometric objects (such as $\sigma^{\alpha'}$) that appear in the 
covariant expressions. I find that two coordinate systems are
particularly useful in this context: the {\it Fermi normal
coordinates} $(t,x^a = s\omega^a)$, and the {\it retarded null-cone
coordinates} $(u,x^a = r\Omega^a)$. 

A third coordinate system, known as the {\it Thorne-Hartle-Zhang
coordinates} \cite{thorne-hartle:85, zhang:86}, has also appeared in
the self-force literature --- they are the favoured choice of the
Florida group led by Steve Detweiler and Bernard Whiting (see, for
example, Ref.~\cite{detweiler:05}). The THZ coordinates are a variant
of the Fermi coordinates, and they have some nice properties. But I
find them less convenient to deal with than the Fermi or retarded
coordinates, because they do not seem to possess a simple covariant
definition. (The THZ coordinates may enjoy the mild Florida winters,
but they are not robust enough to endure the tougher Canadian 
winters.) I shall not discuss the THZ coordinates here.  

\begin{figure} 
\includegraphics[width=0.5\linewidth]{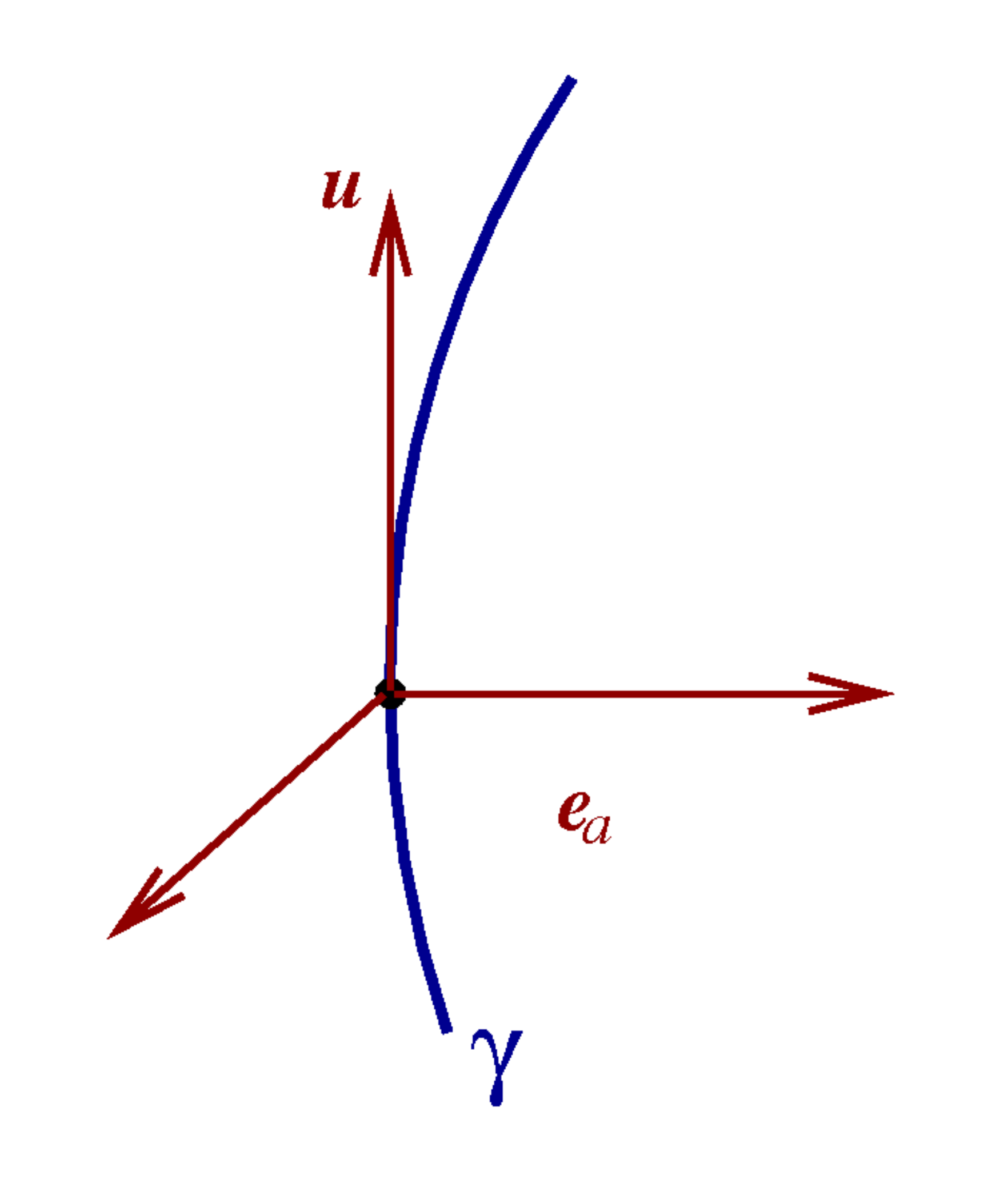}
\caption{A tetrad of basis vectors on the world line $\gamma$. The
  unit timelike vector $u^\mu$ is tangent to the world line. The
  unit spatial vectors $e^\mu_a$ are mutually orthogonal and also
  orthogonal to $u^\mu$.}   
\end{figure} 

The Fermi and retarded coordinates share a basic geometrical
construction on the world line. At each point on $\gamma$ we erect a
basis $(u^\mu, e^\mu_a)$ of orthonormal vectors. The timelike vector
$u^\mu$ is the particle's velocity vector, and it is tangent to the 
world line.  The spatial unit vectors $e^\mu_a$ are labeled with the
index $a = 1, 2, 3$, and they are all orthogonal to $u^\mu$; they are
also mutually orthogonal. The vectors are transported on $\gamma$ so
as to preserve their orthonormality properties. If the world line is a
geodesic, then we might take the vectors $e^\mu_a$ to be parallel
transported on $\gamma$. If instead the world line is accelerated,
then we might take the spatial vectors to be Fermi-Walker transported
on the world line (LRR Sec.~3.2.1). The tetrad of basis vectors
satisfies the completeness relation
\begin{equation} 
g^{\mu\nu} = -u^{\mu} u^{\nu} + \delta^{ab} e^\mu_a e^\nu_b, 
\end{equation}
which holds at any point on the world line. 

Any tensor that is evaluated on $\gamma$ can be decomposed in the
basis $(u^\mu, e^\mu_a)$. For example, we might introduce the 
{\it frame components} of the Riemann tensor, 
\begin{subequations}
\begin{align} 
R_{0a0b}(\tau) &:=  R_{\mu\alpha\nu\beta} \Bigr|_\gamma
u^\mu e^\alpha_a u^\nu e^\beta_b, \\ 
R_{0abc}(\tau) &:=  R_{\mu\alpha\beta\gamma} \Bigr|_\gamma
u^\mu e^\alpha_a e^\beta_b e^\gamma_c, \\ 
R_{abcd}(\tau) &:=  R_{\alpha\beta\gamma\delta} \Bigr|_\gamma
e^\alpha_a e^\beta_b e^\gamma_c e^\delta_d. 
\end{align}
\end{subequations} 
They are functions of proper time $\tau$ on the world line.  

\begin{figure} 
\includegraphics[width=0.6\linewidth]{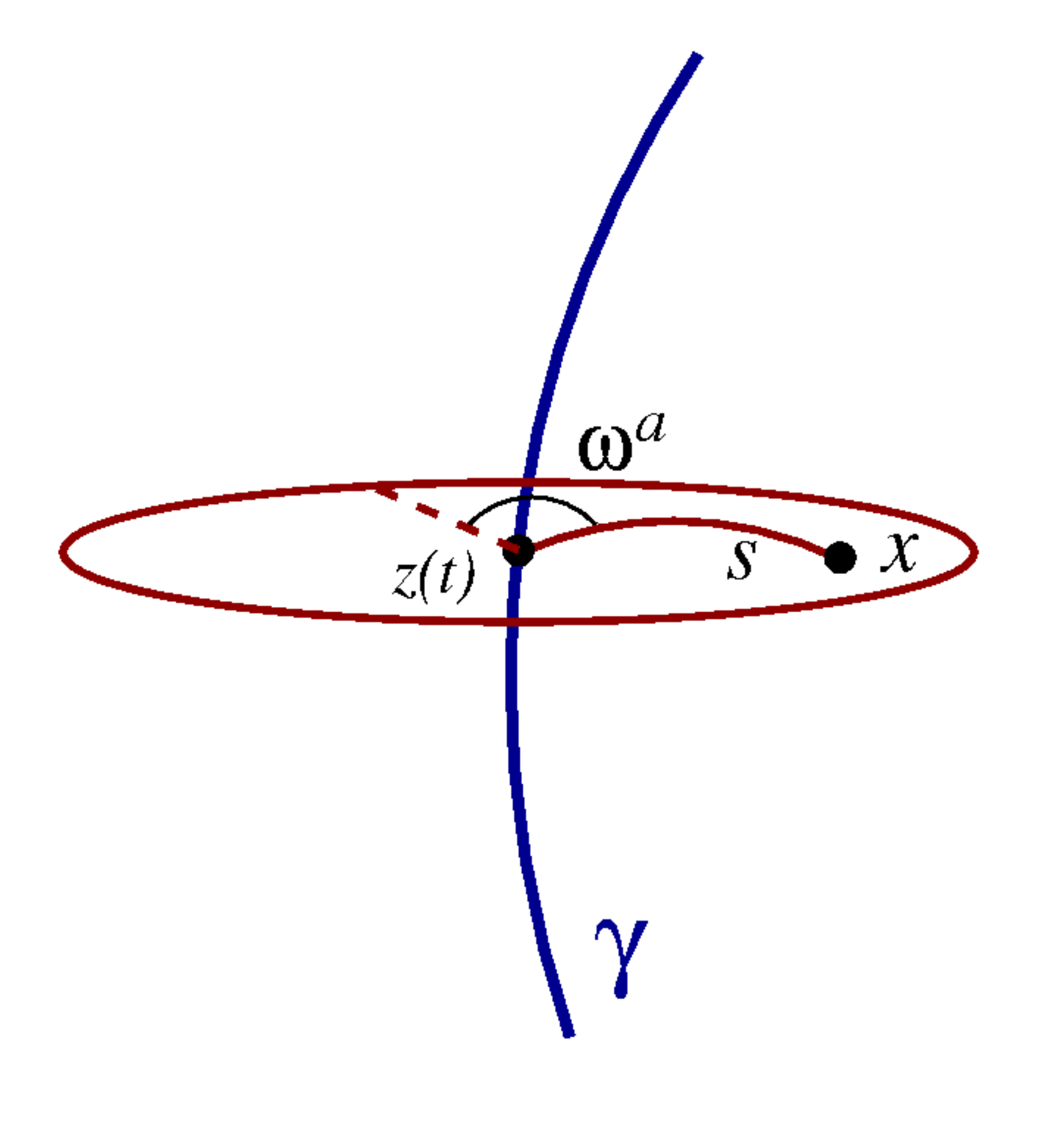}
\caption{Fermi coordinates.}   
\end{figure} 

The Fermi coordinates $(t, x^a = s\omega^a)$ are constructed as
follows (LRR Sec.~3.2). We select a point $x$ in a neighbourhood of
the world line, and we locate the unique geodesic segment $\beta$ that
originates at $x$ and intersects $\gamma$ orthogonally. The
intersection point is labeled $\bar{x}$, and $t$ is the value
of the proper-time parameter at this point: $\bar{x} = z(\tau =
t)$. This defines the time coordinate $t$ of the point $x$. The
spatial coordinates $x^a$ are defined by 
\begin{equation} 
x^a := - e^a_{\bar{\alpha}}(\bar{x}) \sigma^{\bar{\alpha}}(x,\bar{x}); 
\end{equation} 
they are the projections in the basis $e^{\bar{\alpha}}_a$ of the
deviation vector $-\sigma^{\bar{\alpha}}(x,\bar{x})$ between the
points $x$ and $\bar{x}$. The Fermi coordinates come with the
condition  $\sigma_{\bar{\alpha}}(x,\bar{x}) u^{\bar{\alpha}}(\bar{x})
= 0$, which states that the deviation vector is orthogonal to the
world line's tangent vector; it is this condition that identifies the
intersection point $\bar{x} = z(t)$. 

It is useful to introduce $s$ as the proper distance between $x$ and 
the world line. This is formally defined by $s^2 :=
2\sigma(x,\bar{x})$, and it is easy to involve the completeness
relation and show that $s^2 = \delta_{ab} x^a x^b$ (LRR Sec.~3.2.3);
the Fermi distance $s$ is therefore the usual Euclidean distance
associated with the quasi-Cartesian coordinates $x^a$. It is also
useful to introduce the direction cosines $\omega^a
:= x^a/s$; these quantities satisfy $\delta_{ab} \omega^a \omega^b =
1$, and they can be thought of as a radial unit vector that points
away from the world line. 

Each hypersurface $t = \mbox{constant}$ is orthogonal to the world  
line (in the sense described above), and each spacetime point $x$
within the surface can be said to be simultaneous with $\bar{x} =
z(t)$. The Fermi coordinates therefore provide a convenient notion of
{\it rest frame} for the particle.  

\begin{figure} 
\includegraphics[width=0.6\linewidth]{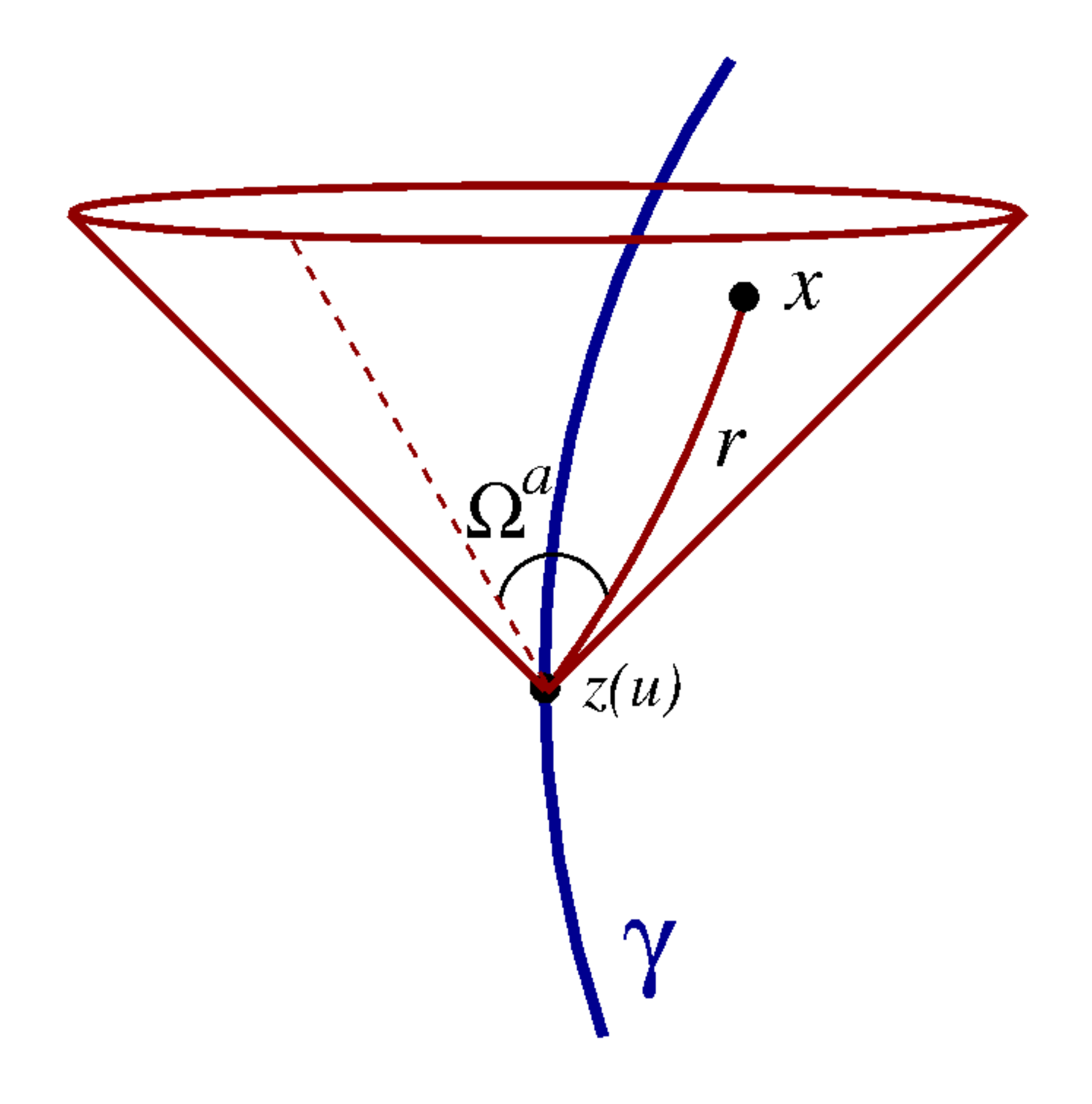}
\caption{Retarded coordinates.}   
\end{figure} 

The retarded coordinates $(u, x^a = r\Omega^a)$ are constructed as 
follows (LRR Sec.~3.3). Once more we select a point $x$ in a
neighbourhood of the world line, but this time we locate the unique
{\it null geodesic segment} $\beta$ that originates at $x$ and travels
backward in time toward $\gamma$. The new intersection point is
labeled $x'$, and $u$ is the value of the proper-time parameter at
this point: $x' = z(\tau = u)$. This defines the time coordinate $u$
of the point $x$. The spatial coordinates $x^a$ are defined exactly as
before, by   
\begin{equation} 
x^a := - e^a_{\alpha'}(x') \sigma^{\alpha'}(x,x'). 
\end{equation} 
The retarded coordinates come with the condition  $\sigma(x,x') = 0$,
which states that the points $x$ and $x' = z(u)$ are linked by a null
geodesic (which travels forward in time from $x'$ to $x$).  

It is useful to introduce $r$ as a measure of light-cone distance
between $x$ and the world line. This is formally defined by $r :=
\sigma_{\alpha'}(x,x') u^{\alpha'}(x')$, and can be shown to be an
affine parameter on the null geodesic that links $x$ to $x'$ (LRR
Sec.~3.3.3). In addition, we have that $r^2 = \delta_{ab} x^a x^b$,
and $r$ is the usual Euclidean distance associated with the spatial 
coordinates $x^a$. It is also useful to introduce the direction
cosines $\Omega^a := x^a/r$, which again play the role of a unit
radial vector that points away from the world line. 

Each hypersurface $u = \mbox{constant}$ is the future light cone of 
the point $z(u)$ on the world line. Any point $x$ on this light cone is
in direct causal contact with $z(u)$. For this reason the retarded
coordinates give the simplest description of the scalar field $\Phi$
produced by a point charge $q$ moving on the world line. The field
satisfies a wave equation, and the radiation produced by the field
essentially realizes the light cones that are so prominently featured
in the construction of the coordinates.  

\section{Field equation and particle motion} 

Let me recapitulate the problem that we wish to solve. We have a point 
particle of mass $m$ and scalar charge $q$ moving on a world line
$\gamma$ described by the parametric relations $z^\mu(\tau)$. The
particle creates a scalar field $\Phi(x)$ and this field acts back on
the particle and produces a force $F_{\rm self}^\alpha$. We wish to
determine this self-force. 

The scalar field obeys the linear wave equation 
\begin{equation} 
\Box \Phi = -4\pi \mu,  
\end{equation} 
where $\Box := g^{\alpha\beta} \nabla_\alpha \nabla_\beta$ is the wave
operator, and 
\begin{equation} 
\mu(x) = q \int_\gamma \delta_4\bigl(x,z(\tau) \bigr)\, d\tau
\end{equation} 
is the scalar-charge density, expressed as an integral over the world
line. The four-dimensional delta function $\delta_4(x,z)$ is defined
as a scalar quantity; it is normalized by 
$\int \delta_4(x,x')\, \sqrt{-g} d^4x  = 1$. 

The particle moves according to 
\begin{equation}  
m(\tau) a^\mu = q \bigl( g^{\mu\nu} + u^\mu u^\nu\bigr) \Phi_\nu, 
\label{EoM}
\end{equation} 
where $a^\mu = Du^\mu/d\tau$ is the covariant acceleration and
$\Phi_\nu := \nabla_\nu \Phi$ is the field gradient. The presence of
the projector $g^{\mu\nu} + u^\mu u^\nu$ on the right-hand side
ensures that the acceleration is orthogonal to the velocity. The field
gradient, however, also has a component in the direction of the world
line, and this produces a change in the particle's rest mass (LRR
Sec.~5.1.1): $dm/d\tau = -q u^\mu \Phi_\mu$. We shall not be concerned 
with this effect here. Suffice it to say that the mass is not
conserved because the scalar field can radiate monopole waves, which
is impossible for electromagnetic and gravitational radiation. 

The wave equation for $\Phi$ can be integrated, as we shall do in the
following section, and the solution can be examined near the world
line. Not surprisingly, the field is singular on the world line. This
property makes the equation of motion meaningless as it stands, and we
shall have to make sense of it in the course of our analysis. 

\section{Retarded Green's function} 

The wave equation for $\Phi$ is solved by means of a {\it Green's
function} $G(x,x')$ that satisfies  
\begin{equation} 
\Box G(x,x') = -4\pi \delta_4(x,x').  
\end{equation} 
The solution is simply 
\begin{equation} 
\Phi(x) = \int G(x,x') \mu(x')\, \sqrt{-g'} d^4x' 
= q \int_\gamma G(x,z)\, d\tau, 
\end{equation} 
and the difficulty of solving the wave equation has been transferred
to the difficulty of computing the Green's function. We wish to
construct the {\it retarded solution} to the wave equation, and this
is accomplished by selecting the retarded Green's function 
$G_{\rm ret}(x,x')$ among all the solutions to Green's
equation. (Other choices will be considered below.) The retarded
Green's function possesses the important property that it vanishes
when the source point $x'$ is in the future of the field point
$x$. This ensures that $\Phi(x)$ depends on the past behaviour of the
source $\mu$, but not on its future behaviour.  

The retarded Green's function is known to exist globally as a
distribution if the spacetime is globally hyperbolic. But knowledge of
the Green's function is required only in the immediate vicinity of the
world line, so as to identify the behaviour of $\Phi$ there; we shall
not be concerned with the behaviour of the Green's function when $x$
and $z(\tau)$ are widely separated.  

In this context the Green's function can be shown (LRR Sec.~4.3) to
admit a {\it Hadamard decomposition} of the form 
\begin{equation} 
G_{\rm ret}(x,x') = U(x,x') \delta_{\rm future}(\sigma) 
+ V(x,x') \Theta_{\rm future}(-\sigma), 
\end{equation} 
where $\sigma(x,x')$ is the world function introduced previously, and
the two-point functions $U(x,x')$ and $V(x,x')$ are smooth when
$x \to x'$. The retarded Green's function is not smooth in this limit,
however, as we can see from the presence of the delta and theta
functions. The first term involves $\delta_{\rm future}(\sigma)$, the
restriction of $\delta(\sigma)$ on the future light cone of the source
point $x'$.  The delta function is active when $\sigma(x,x') = 0$, and
this describes (for fixed $x'$) the future and past light cones of
$x'$. We then eliminate the past branch of the light cone --- for
example, by multiplying $\delta(\sigma)$ by the step function
$\Theta(t-t')$ --- and this produces 
$\delta_{\rm future}(\sigma)$. The second term involves 
$\Theta_{\rm future}(-\sigma)$, a step function that is
active when $\sigma < 0$, that is, when $x$ and $x'$ are timelike
related; we also restrict the interior of the light cone to the future
branch, so that $x$ is necessarily in the future of $x'$. 

The delta term in $G_{\rm ret}(x,x')$ is sometimes called the 
{\it direct term}, and it corresponds to propagation from $x'$ to $x$
that takes place directly on the light cone. If the Green's function
contained a direct term only (as it does in flat spacetime), the field
at $x$ would depend only on the conditions of the source $\mu$ at the 
corresponding retarded events $x'$, the intersection between the
support of the source and $x$'s past light cone. In the case of a
point particle this reduces to a single point $x' \equiv z(u)$. The
theta term in $G_{\rm ret}(x,x')$, which is sometimes called the 
{\it tail term}, corresponds to propagation within the light cone;
this extra term (which is generically present in curved spacetime) 
brings a dependence from events $x'$ that lie in the past of the
retarded events. In the case of a point particle, the field at $x$
depends on the particle's entire past history, from $\tau = -\infty$ to
$\tau = u$. 

There exists an algorithm to calculate $U(x,x')$ and $V(x,x')$ in the
form of Taylor expansions in powers of $-\sigma^{\alpha'}$
(LRR Sec.~4.3.2). It returns
\begin{equation} 
U(x,x') = 1 + \frac{1}{12} R_{\alpha'\beta'} \sigma^{\alpha'}
\sigma^{\beta'} + \cdots 
\end{equation}
and 
\begin{equation} 
V(x,x') = \frac{1}{12} R(x') + \cdots, 
\end{equation} 
where $R_{\alpha'\beta'}$ is the Ricci tensor at $x'$, and $R(x')$
the Ricci scalar at $x'$. In Ricci-flat spacetimes the expansions of
$U - 1$ and $V$ both begin at the fourth order in $\sigma^{\alpha'}$.  

\section{Alternate Green's function} 

The {\it advanced} Green's function is given by (LRR Sec.~4.3)
\begin{equation}
G_{\rm adv}(x,x') = U(x,x') \delta_{\rm past}(\sigma) 
+ V(x,x') \Theta_{\rm past}(-\sigma), 
\end{equation} 
in terms of the same two-point functions $U(x,x')$ and $V(x,x')$ that
appear within the retarded Green's function. The difference is that
the light cones are now restricted to the past branch, so that 
$G_{\rm adv}(x,x')$ vanishes when $x'$ is in the past of $x$.  A
solution to the wave equation constructed with the advanced Green's
function would display anti-causal behaviour: it would depend on the
future history of the source. 

Another useful choice of Green's function is the 
{\it Detweiler-Whiting singular} Green's function defined by (LRR
Sec.~4.3.5) 
\begin{equation}  
G_{\rm S}(x,x') = \frac{1}{2} U(x,x') \delta(\sigma) 
- \frac{1}{2} V(x,x') \Theta(\sigma). 
\end{equation} 
Here the delta and theta functions are no longer restricted: both
future and past branches contribute to the Green's function. In fact,
the argument of the step function is now $+\sigma$, and this indicates
that the second term is active when $x$ and $x'$ are spacelike
related. For fixed $x'$, the singular Green's function is nonzero when
$x$ is either on, or outside, the (past and future) light cone of
$x'$. Unlike the retarded and advanced Green's functions, 
$G_{\rm S}(x,x')$ is not known to exist globally as a distribution 
(even for globally hyperbolic spacetimes); its local existence is not
in doubt, however, and this suffices for our purposes.

\begin{figure} 
\includegraphics[angle=-90, width=0.5\linewidth]{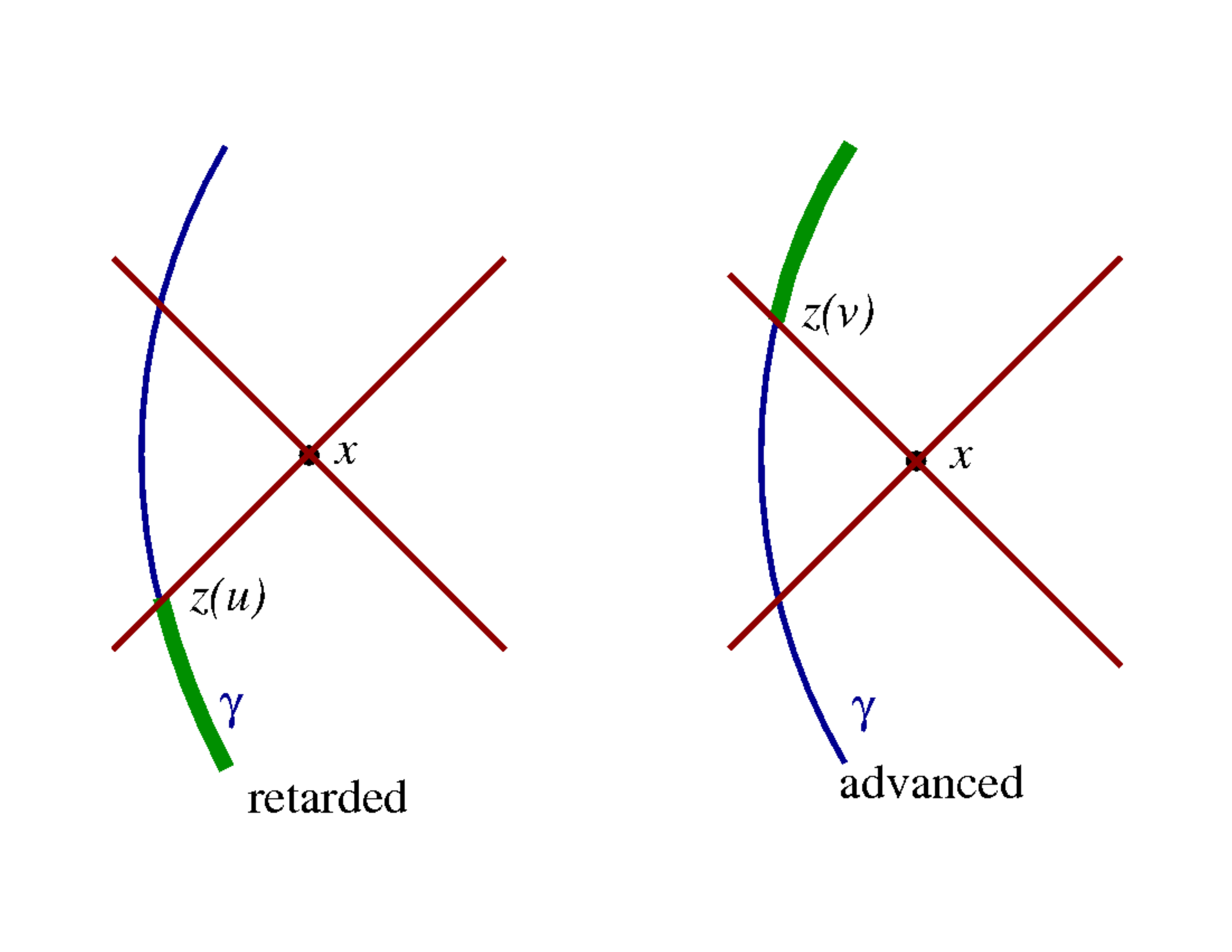}
\hfill
\includegraphics[angle=-90, width=0.5\linewidth]{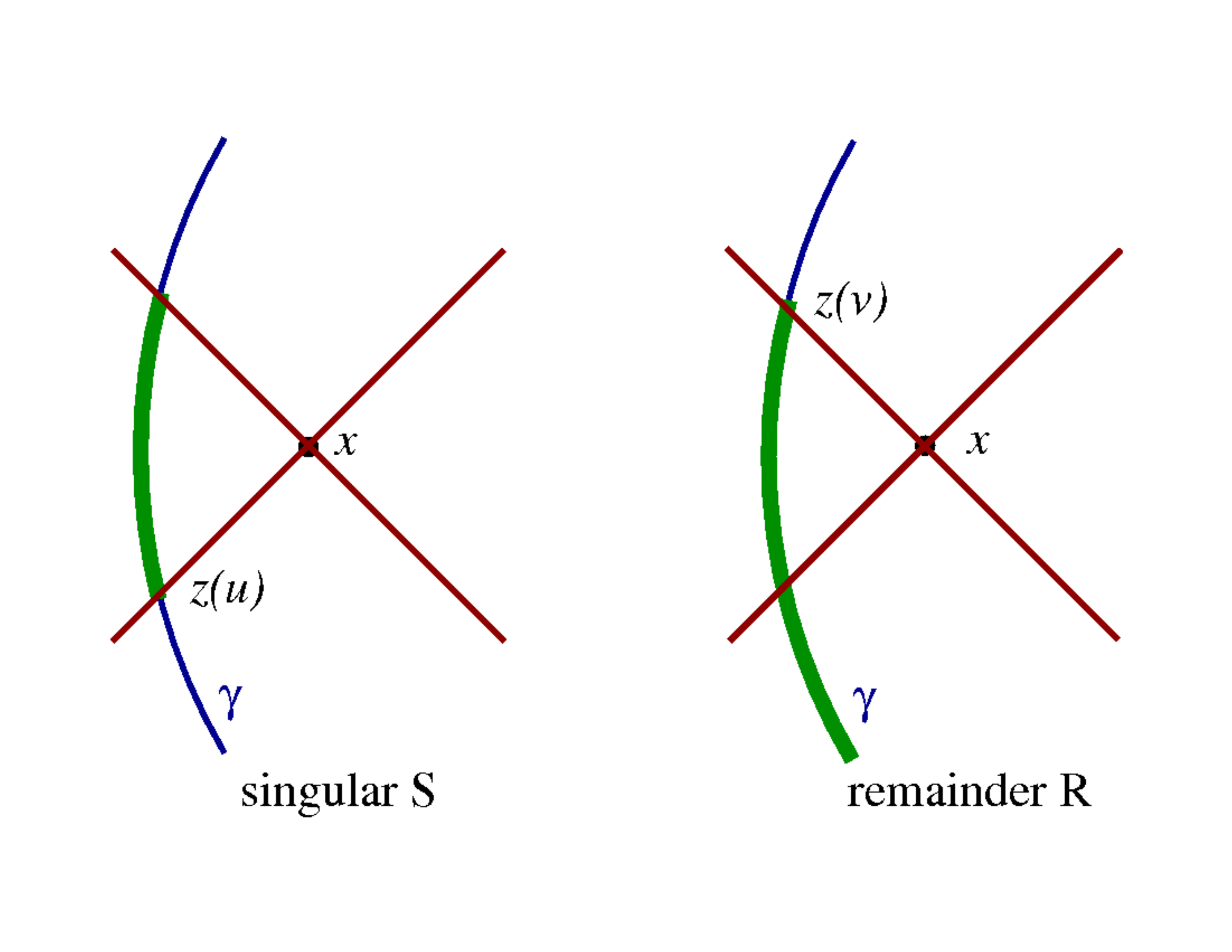}
\caption{Solutions to the wave equation. The retarded field is
  sourced by the past history of the point particle, up to the
  retarded position $z(u)$. The advanced field is sourced by the
  future history, starting from the advanced position $z(v)$. The
  singular field is sourced by the world-line segment that lies on and 
  outside the light cone of the field-point $x$, starting from the
  retarded position $z(u)$ and ending at the advanced position
  $z(v)$. Finally, the regular field is sourced by the history of the
  particle up to the advanced position $z(v)$.}   
\end{figure} 

All three Green's functions satisfy the same wave equation, $\Box
G = -4\pi \delta_4$, and all three give rise to fields $\Phi(x)$ that
diverge on the particle's world line. A useful combination of Green's
functions is the {\it Detweiler-Whiting regular} two-point function  
\begin{equation} 
G_{\rm R}(x,x') = G_{\rm ret}(x,x') - G_{\rm S}(x,x'),
\end{equation} 
which satisfies the homogeneous wave equation $\Box G_{\rm R}(x,x') 
= 0$. This two-point function gives rise to a field 
\begin{equation} 
\Phi_{\rm R} = \Phi_{\rm ret} - \Phi_{\rm S} 
\end{equation} 
that also satisfies the homogeneous wave equation, 
$\Box \Phi_{\rm R} = 0$. This field is the difference between two
singular fields, and since $\Phi_{\rm ret}$ and $\Phi_{\rm S}$ are
equally singular near the world line, we find that the {\it regular
remainder field} $\Phi_{\rm R}$ stays bounded when $x \to
z(\tau)$. And what's more, the regular field is smooth, in the sense
that it and any number of its derivatives possess a well-defined limit
when $x \to z(\tau)$.   

\section{Fields near the world line} 

Using the ingredients presented in the preceding sections it is
possible to show (LRR Secs.~5.1.3, 5.1.4, and 5.1.5) that close to the
world line, the retarded, singular, and regular fields are given by 
\begin{subequations} 
\begin{align} 
\Phi_{\rm ret} &= \frac{q}{r} + q \int_{-\infty}^u V(x,z)\, d\tau 
+ O(r^2), \\ 
\Phi_{\rm S} &= \frac{q}{r} - \frac{1}{3} q \dot{a}_a x^a  
+ O(r^2), \\ 
\Phi_{\rm R} &= \frac{1}{3} q \dot{a}_a x^a
+ q \int_{-\infty}^u V(x,z)\, d\tau + O(r^2), 
\end{align}
\end{subequations} 
in the retarded coordinates $(u,x^a = r\Omega^a)$. These expressions
involve the retarded distance $r$ from $x$ to $z(u)$ and the frame
component $\dot{a}_a(u) := \dot{a}_\mu e^\mu_a$ of 
$\dot{a}^\mu := D a^\mu/d\tau$, the proper-time derivative of the
acceleration vector. They involve also an integration over the past
history of the particle. These expressions are valid for Ricci-flat
spacetimes.  We observe that the retarded and singular fields both
diverge as $q/r$ as we approach the world line, but that the regular
remainder $\Phi_{\rm R}$ is free of singularities. 

In the Fermi coordinates $(t, x^a = s\omega^a)$ we have the more
complicated expressions 
\begin{subequations} 
\begin{align} 
\Phi_{\rm ret} &= \frac{q}{s} - \frac{1}{2} q a_a \omega^a 
+ qs \biggl( \frac{1}{8} \dot{a}_0 + \frac{1}{3} \dot{a}_a \omega^a 
- \frac{1}{6} R_{0a0b} \omega^a \omega^b \biggr) 
\nonumber \\  & \qquad \mbox{} 
+ q \int_{-\infty}^{t} V(x,z)\, d\tau + O(s^2), \\ 
\Phi_{\rm S} &= \frac{q}{s} - \frac{1}{2} q a_a \omega^a 
+ qs \biggl( \frac{1}{8} \dot{a}_0 
- \frac{1}{6} R_{0a0b} \omega^a \omega^b \biggr) + O(s^2), \\ 
\Phi_{\rm R} &= \frac{1}{3} q \dot{a}_a x^a
+ q \int_{-\infty}^{t} V(x,z)\, d\tau + O(s^2).
\end{align}
\end{subequations} 
They involve the spatial distance $s$ between $x$ and $z(t)$, the
frame components of the acceleration vector and its proper-time
derivative, the frame components of the Riemann tensor, and the
integration over the past history of the particle. Once more we see
that the retarded and singular fields diverge in the limit $s \to 0$,
but that the regular field is free of singularities. (The singular
nature of $\Phi_{\rm S}$ is also observed in a term such as
$\frac{1}{2} q a_a \omega^a$ that stays bounded when $s \to 0$, but is
directionally ambiguous on the world line.) 

From these equations we may calculate the spatial derivatives of the
singular and regular fields. We obtain  
\begin{subequations} 
\label{gradients} 
\begin{align} 
\nabla_a \Phi_{\rm S} &= -\frac{q}{s^2} \omega_a 
- \frac{q}{2s} \bigl( \delta^b_a - \omega^b \omega_a \bigr) a_b 
+ \frac{q}{8} \dot{a}_0\, \omega_a 
\nonumber \\ & \qquad \mbox{} 
+ \frac{q}{6} R_{0b0c}\, \omega_a \omega^b \omega^c 
- \frac{q}{3} R_{0a0b}\, \omega^b + O(s), \\ 
\nabla_a \Phi_{\rm R} &=\frac{1}{3} q \dot{a}_a 
+ q \int_{-\infty}^t \nabla_a V(x,z)\, d\tau + O(s).  
\end{align}
\end{subequations} 
As expected, the gradient of the singular field diverges as $q/s^{2}$,
but the gradient of  the regular field is free of singularities. The
gradient of the retarded field is $\nabla_a \Phi_{\rm ret} = 
\nabla_a \Phi_{\rm S} + \nabla_a \Phi_{\rm R}$. 
  
We notice that many terms in $\nabla_a \Phi_{\rm S}$ are proportional
to an odd number of radial vectors $\omega^a$; all such terms vanish
when we average the field gradient over a spherical surface of
constant $s$. This averaging leaves behind   
\begin{equation} 
\bigl\langle \nabla_a \Phi_{\rm S} \bigr\rangle 
= -\biggl( \frac{q}{3s} \biggr) a_a+ O(s), 
\end{equation} 
which is still singular. To obtain this result we made use of the
identity $\langle \omega^b \omega_a \rangle = \frac{1}{3}
\delta^b_a$. Notice that because $\nabla_a \Phi_{\rm R}$ is
smooth at $s=0$, an averaging simply returns the same expression: 
$\langle \nabla_a \Phi_{\rm R} \rangle = \nabla_a \Phi_{\rm R}$.

\section{Self-force} 

Let us now reflect on the results of the preceding section and try to 
make sense of Eq.~(\ref{EoM}) as an equation of motion for the
particle. Our first attempt will be entirely heuristic; we shall add
refinement to our treatment in the following section.  

Given that Eq.~(\ref{EoM}) does not make sense as it stands when we 
insert $\Phi = \Phi_{\rm ret}$ on its right-hand side, let us take
the view that the equation is meant to apply to an extended body
instead of a point particle, and let us average $\Phi_\mu :=
\nabla_\mu \Phi$ over the body's volume. This operation should be
carried out in the body's rest frame, and for this purpose it is
natural to adopt the Fermi coordinates. We aim, therefore, to average 
the spatial components $\Phi^{\rm ret}_a$ of the field gradient. The
simplest form of averaging was carried out already in the preceding
section, and we obtained  
\begin{equation} 
\langle \Phi^{\rm ret}_a \rangle = 
\langle \Phi^{\rm S}_a \rangle + \langle \Phi^{\rm R}_a \rangle 
= -\biggl( \frac{q}{3s} \biggr) a_a + \Phi^{\rm R}_a+ O(s), 
\end{equation} 
in which $\Phi^{\rm R}_a$ is evaluated (without obstacle) at
$s=0$. This expression corresponds to pretending that the body is a 
thin spherical shell of radius $s$.  

Substitution into Eq.~(\ref{EoM}) and evaluation in the Fermi
coordinates produces 
\begin{equation} 
( m + \delta m ) a^a = q  \Phi_{\rm R}^a, 
\end{equation} 
with $\delta m := q^2/(3s)$ denoting the contribution to the
total body mass that comes from the field's energy. Absorbing this
into a redefinition of the inertial mass $m$, the final tensorial
expression for the equation of motion is 
\begin{equation} 
m a^\alpha = q \bigl( g^{\alpha\beta} + u^\alpha u^\beta \bigr) 
\nabla_\beta \Phi_{\rm R}, 
\label{Quinn} 
\end{equation} 
with 
\begin{equation} 
\nabla_\beta \Phi_{\rm R} = \frac{1}{3} q \dot{a}_\beta 
+ q \int_{-\infty}^\tau \nabla_\beta V(x,z)\, d\tau'. 
\end{equation} 
This is Quinn's equation of motion \cite{quinn:00} for a scalar charge
$q$ moving in a curved spacetime with metric $g_{\alpha\beta}$. The
self-force involves an instantaneous term proportional to
$\dot{a}^\alpha = D a^\alpha/d\tau$, as well as an integral over the
particle's past history. 

Equation (\ref{Quinn}) informs us that of the complete retarded field
$\Phi_{\rm ret} = \Phi_{\rm S} + \Phi_{\rm R}$, only the
Detweiler-Whiting regular field $\Phi_{\rm R}$ contributes to the
self-force. The role of the singular field is merely to contribute to
the particle's inertia, through a shift $\delta m$ in its inertial
mass. This contribution diverges in the limit $s \to 0$, but it would
be finite for any extended body.  

I must confess that this computation returns the wrong expression for
the particle's self-energy. We obtained $\delta m = q^2/(3s)$, while
the correct expression is $\delta m = q^2/(2s)$; we are wrong by a
factor of $2/3$.  I believe that this discrepancy originates in an
inconsistency between our assumed shape for the extended body --- a
spherical shell of radius $s$ ---  and the field it produces, which we
took to be equal to the field produced by a point
particle. I would conjecture that calculating the field actually
produced by a spherical shell would give rise to the correct
expression for $\delta m$, but leave unchanged the final result of
Eq.~(\ref{Quinn}) for the equation of motion. 

\section{Axiomatic approach} 

The procedure outlined above is admittedly heuristic. It can, however,
be formalized and put on an axiomatic basis that supplies
Eq.~(\ref{Quinn}) with a much improved pedigree. This is the approach
that was first pursued by Ted Quinn and Bob Wald \cite{quinn-wald:97,
  quinn:00}. They formulate two axioms that the scalar self-force
$F^\alpha$ should satisfy: 

\begin{description} 
\item[{\bf Quinn-Wald Axiom 1.}] Two scalar particles move on world
lines $\gamma$ and  $\tilde{\gamma}$ in two different spacetimes. At
points $z$ and $\tilde{z}$ their acceleration vectors have equal
lengths. The neighbourhoods of $z$ and $\tilde{z}$, as well as the
acceleration vectors, are identified in Fermi coordinates. Then the
difference in the self-forces is given by   
\begin{equation} 
F^\alpha - \tilde{F}^\alpha = 
q \bigl( g^{\alpha\beta} + u^\alpha u^\beta \bigr) 
\lim_{s \to 0} \Bigl\langle \Phi_\beta  - \tilde{\Phi}_\beta
\Bigr\rangle.  
\end{equation} 
Here $\Phi$ and $\tilde{\Phi}$ are the retarded fields in each
spacetime, and $\Phi_\beta := \nabla_\beta \Phi$ while
$\tilde{\Phi}_\beta := \tilde{\nabla}_\beta \tilde{\Phi}$. The limit
is well defined after the difference of field gradients is averaged
over a sphere of radius $s$.   
\item[{\bf Quinn-Wald Axiom 2.}] $\tilde{F}^\alpha = 0$ in flat spacetime, for 
a particle with uniform acceleration.  
\end{description}
The first axiom is essentially a statement that when two particles
momentarily share the same acceleration, their fields are equally
singular, and the difference (after averaging) possesses a
well-defined limit when $s \to 0$. The second axion is a scalar-charge
analogue to a well-known result from flat-spacetime electrodynamics: a
charged particle moving with a uniform acceleration does not undergo
radiation reaction. 

According to Eqs.~(\ref{gradients}), the curved-spacetime expression
for the gradient of the retarded field is 
\begin{eqnarray} 
\Phi_a &=& -\frac{q}{s^2} \omega_a 
- \frac{q}{2s} \bigl( \delta^b_a - \omega^b \omega_a \bigr) a_b 
+ \frac{q}{8} \dot{a}_0\, \omega_a 
\nonumber \\ & & \mbox{} 
+ \frac{q}{6} R_{0b0c}\, \omega_a \omega^b \omega^c 
- \frac{q}{3} R_{0a0b}\, \omega^b + \nabla_a \Phi_{\rm R} + O(s); 
\end{eqnarray} 
this holds at time $t$ in Fermi coordinates. The flat-spacetime
expression is  
\begin{equation}
\tilde{\Phi}_a = -\frac{q}{s^2} \omega_a 
- \frac{q}{2s} \bigl( \delta^b_a - \omega^b \omega_a \bigr) a_b 
+ O(s), 
\end{equation} 
and this also holds at time $t$ in the same system of Fermi
coordinates. Under the conditions of the Quinn-Wald axioms, the
acceleration $a_a$ that appears in $\Phi_a$ and $\tilde{\Phi}_a$ is
one and the same. In the flat-spacetime expression we set $\dot{a}_0$
and $\dot{a}_a$ to zero because the acceleration is chosen to be
uniform. In addition we eliminate the Riemann-tensor terms,  as well
as the integral over the particle's past history --- $V$ necessarily
vanishes in flat spacetime. 

Subtraction yields 
\begin{equation} 
\Phi_a - \tilde{\Phi}_a =  \frac{q}{8} \dot{a}_0\, \omega_a 
+ \frac{q}{6} R_{0b0c}\, \omega_a \omega^b \omega^c 
- \frac{q}{3} R_{0a0b}\, \omega^b + \nabla_a \Phi_{\rm R} + O(s), 
\end{equation}
and we get 
\begin{equation} 
\bigl\langle \Phi_a - \tilde{\Phi}_a \bigr\rangle 
=  \nabla_a \Phi_{\rm R} + O(s)
\end{equation}
after averaging over a sphere of constant $s$. The difference in the
self-forces is therefore $F_a - \tilde{F}_a =  q\nabla_a 
\Phi_{\rm R}$. The second axiom finally returns $F_a = q \nabla_a 
\Phi_{\rm R}$, which is equivalent to Eq.~(\ref{Quinn}); 
we have reproduced Quinn's expression for the scalar self-force.
Notice that the second axiom eliminates the need to carry out an
explicit renormalization of the mass.  
  
Another axiomatic approach provides an even more immediate derivation
of Quinn's equation. This is the approach suggested by Steve Detweiler
and Bernard Whiting  \cite{detweiler-whiting:03}, which is based on an
observation and an alternate axiom:  
\begin{description} 
\item[{\bf Detweiler-Whiting Observation.}] The retarded field
$\Phi_{\rm ret}$ can be decomposed uniquely into a singular piece
$\Phi_{\rm S}$ and a regular remainder $\Phi_{\rm R}$. 
\item[{\bf Detweiler-Whiting Axiom.}] The singular field produces no
force on the particle.
\end{description} 
The immediate consequence of the axiom is that only the regular field
participates in the self-force, and we once more arrive at Quinn's
equation.  

The Detweiler-Whiting approach is very clean and provides a quick
route to the final answer. The observation is not at all
controversial, because the singular field is indeed uniquely defined
by the prescription outlined in Sec.~6. The axiom, on the other hand, 
seems too good to be true. How can it just be asserted that the 
singular field produces no force? 

A fairly compelling line of argument rests on the fact that according
to its definition, the singular field is strongly time-symmetric, in
the sense that the field at $x$ does not depend on the future nor the
past of the spacetime point; it instead depends on source points $x'$
that are in a spacelike or lightlike relation with $x$. Since we would
expect the self-force to be sensitive to the direction of time --- an
advanced field should produce a different force  from a retarded field
--- it seems plausible that the singular field would not know whether
to push or pull, and would therefore choose to do neither. 

The argument is not water-tight. For example, an alternate 
singular field, defined by Dirac's prescription $\frac{1}{2} 
\Phi_{\rm ret} + \frac{1}{2} \Phi_{\rm adv}$, would also be
time-symmetric (though not strongly time-symmetric), and could also be
asserted to produce no force. The resulting self-force, however, would
be produced by $\frac{1}{2} \Phi_{\rm ret} 
- \frac{1}{2} \Phi_{\rm adv}$, and would depend on the entire history
of the particle, both past and future. We would of course reject this
candidate self-force on grounds of causality violation, but the
argument nevertheless shows that there is more to the
Detweiler-Whiting singular field than a time-symmetry
property. Another hole in the argument lies in the link between the  
time-symmetry of the singular field and the statement that it must
exert no force: While the time-symmetry property clearly implies
that the singular field cannot produce dissipative effects on the
particle, there is no reason to rule out an eventual conservative
contribution to the self-force.  

The conclusion is that additional axioms are necessarily required to
make sense of the equations of motion formulated for a point
particle. The axioms may seem plausible and perhaps even self-evident,
but they cannot be derived from first principles in the context of a
classical field theory coupled to a point particle. Such a theory is
inherently singular and ambiguous, and it necessarily requires
external input in the form of additional axioms.   

\section{Conclusion} 

Can one do better than this? The answer is `no' if we insist in
treating the point particle as a fundamental classical object. The
answer, however, is `yes' if we properly understand that a point
particle is merely a convenient substitute for what is fundamentally
an extended body. In this view, the length scale of the moving body is
$\ell$, not zero. The body possesses a finite density of scalar
charge, the scalar field is finite everywhere, and its motion traces a
world tube in spacetime instead of a single world line. To determine
this motion is a well-posed problem, but the  description now involves
a lot of additional details. Under usual circumstances, however,
$\ell$ is much smaller than all other length scales present in the
problem, such as the radius of curvature ${\cal R}$ of the body's
trajectory. Under these circumstances the description of the motion 
can be simplified so as to involve a much smaller number of variables;
in the limit $\ell/{\cal R} \to 0$ only the position of the
center-of-mass matters, and all couplings between the body's multipole
moments and the external field become irrelevant. In this limit we
recover a point-particle description, with the essential understanding
that it is merely an approximate description that should not be
considered to be fundamental.  

To go through the details of this program is difficult, and it appears
that very few authors have attempted it since the old days of Lorentz
and Abraham. For a recent discussion, and a review of this literature,
see the work by Harte \cite{harte:06}. Another important exception
concerns the gravitational self-force acting on a small black hole
(LRR Sec.~5.4), which is decidedly not treated as a point mass.     

It is well known that in general relativity, the motion of gravitating 
bodies is determined, along with the spacetime metric, by the Einstein
field equations; the equations of motion are not separately
imposed. This observation provides a means of deriving the
gravitational self-force without having to rely on the fiction
of a point mass. In the powerful method of {\it matched asymptotic
expansions}, the metric of the small black hole, perturbed by the tidal
gravitational field of the external spacetime, is matched to the metric
of the external spacetime, perturbed by the black hole. The equations
of motion are then recovered by demanding that the metric be a valid
solution to the vacuum field equations. In my opinion, this method
(which was first applied to the gravitational self-force problem by
Mino, Sasaki, and Tanaka \cite{mino-etal:97}) gives what is by far the
most compelling derivation of the gravitational self-force. Indeed,
the method is entirely free of conceptual and technical pitfalls ---
there are no singularities (except deep inside the black hole) and
only retarded fields are employed.    

In this assessment I respectfully disagree with my colleague Bob
Wald, who finds that the method incorporates a number of unjustified
assumptions. I would concede that expositions of the method ---
including my own in LRR --- might not have sufficiently clarified some
of its subtle aspects. But I see this as faulty exposition, not as an
intrinsic difficulty with the method of matched asymptotic
expansions. I refer the reader to the recent work by Sam Gralla and
Bob Wald \cite{gralla-wald:08} for their views on this issue, and
their own approach to the motion of an extended body in general 
relativity.   

The introduction of a point particle in a classical field theory
appears at first sight to be severely misguided. This is all the more
true in a nonlinear theory such as general relativity. The lesson
learned here is that surprisingly often, {\it one can get away with
it}. The derivation of the gravitational self-force based on the
method of matched asymptotic expansions does indeed show that the
result obtained on the basis of a point-particle description can be
reliable, in spite of all its questionable aspects. This is a
remarkable observation, and one that carries a lot of convenience: It
is indeed much easier to implement the point-mass description than to
perform the matching of two metrics in two coordinate systems. The
lesson, of course, carries over to the scalar and electromagnetic
cases.  

\begin{acknowledgement}
I wish to thank the organizers of the school for their kind invitation
to lecture; Orl\'eans in the summer is a very nice place to be. I wish
to thank the participants for many interesting discussions. And
finally, I wish to thank Bernard Whiting for his patience. This work
was supported by the Natural Sciences and Engineering Research Council
of Canada. 
\end{acknowledgement}  

\bibliography{../bib/master}
\end{document}